# Hot carriers generated by plasmons: where are they are generated and where do they go from there?

Jacob B Khurgin

Johns Hopkins University

Baltimore MD 21218 USA

A physically transparent unified theory of optically and plasmon-induced hot carrier generation in metals is developed with all the relevant mechanisms included. Analytical expressions that estimate the carrier generation rates, their locations, energy and direction of motion are obtained. Among four mechanisms considered: interband absorption, phonon and defect assisted absorption, electron-electron scattering assisted absorption, and surface –collision assisted absorption (Landau damping), it is the last one that generates hot carriers which are most useful for practical applications in photo detection and photo catalysis.

## Introduction

Current developments in plasmonics [1-3] have caused renewed interest to the issue of absorption in metals. The fact that metal absorption is important factor limiting factor in every imaginable application of plasmonics is universally known[4], and, in the absence of efficient means of reducing the loss, the emphasis has been shifting towards such applications where absorption in metal is not the main factor (typically sensors [5]). More recently the attention has been drawn to the all-together different class of applications, where the absorption is a positive, rather than deleterious, factor – such as plasmonic Schottky photo-detectors[6-8] and light harvesting[9, 10]. In these applications the incoming photons excite surface plasmons, either propagating or localized, and the rapid decay of plasmons causes excitation of hot carriers which, once they reach the metal surface, can produce a photo-current in photodetectors [6-8], or initiate a chemical reaction in photo catalysis[11]. In order to understand the operation and evaluate the performance of these plasmon-assisted schemes, it is important not to know not only the rate with which plasmons decay but the distribution of energies and momentum of the hot carriers excited in the process.

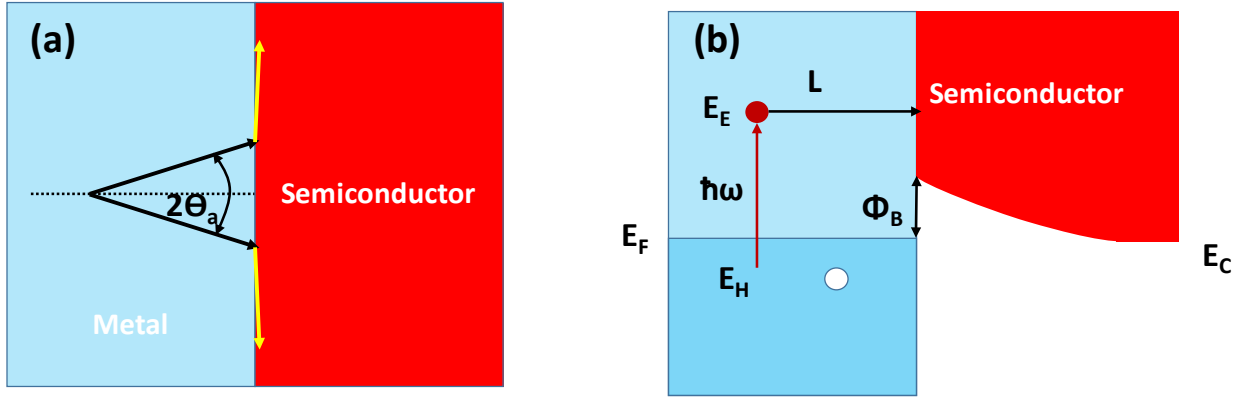

Fig. 1 Hot carrier Schottky barrier metal semiconductor photodetector (a) spatial picture (b) energy band diagram

As a general example consider a plasmonic Schottky detector [6-8] shown in Fig. 1. In this detector decay of a surface plasmon with energy $\hbar\omega$ creates hot electron above the Fermi level with kinetic energy $E_E$ and a hole with energy $E_H$. If the electron has energy its energy above the Schottky barrier $\Phi_B$ it has a chance to be injected into the semiconductor and contribute to photo-current. For that the electron needs to travel unimpeded distance L to the interface, and, more significantly it should have sufficiently small in-plane component of its momentum so it can enter the semiconductor rather than being reflected in manner similar to total internal reflection in optics. This fact has to do with the enormous momentum mismatch between the metal where the carriers have high kinetic energy commensurate with Fermi energy $E_F$, i.e. 5-6 eV in Au or Ag, and semiconductor where the kinetic energy of carriers is only $E_E - \Phi_B$., or at best 1eV. To avoid total reflection, the electrons must propagate inside a cone subtended by an "acceptance angle" [12]

$$\Theta_a = \sin^{-1}\left(\sqrt{\frac{m_s}{m_0}\frac{E_E - \Phi_b}{E_F}}\right) \qquad (1)$$

where $m_s$ and $m_0$ are the effective masses in the semiconductor an metal respectively and is typically on the scale of 10-20 degrees leading to the injection efficiencies often less than 1% for the carriers propagating omnidirectionally. For this reason, it is important to investigate what are the energies as well as spatial and angular distributions of hot carriers engendered by the decay of surface plasmons.

While there has been a significant amount of work performed in this respect recently [13-18], there is no clear consensus achieved and no unified picture of the process has been presented. Four different mechanisms of plasmon decay had been identified – interband absorption, phonon (or defect) assisted absorption, electron-electron scattering assisted absorption, and interband absorption, and each one had been treated using different formalism, ranging from simple phenomenological treatment to a full quantum one. Some of the processes, such as electron-electron scattering assisted absorption had been totally neglected, which, as we show below is not justified in the visible and even IR range. On the other hand, the interband absorption was considered as a promising source of hot carriers, while in reality most of the energy in this process goes to the holes residing in the d-band and having very low velocity which prevents them from reaching the interface and performing any useful function. Furthermore, while most of the studies had been focused on the energy distribution of the hot carriers, little attention [15]had been given to their angular and spatial distributions, which as we have just explained are of paramount importance to hot-carrier enabled devices. Most and foremost, practically all of the current work has been performed numerically which may be a convenient method but it does not reveal the intricate physics of hot carrier generation.

In this work we develop a simple analytical model of hot carrier generation based on Fermi golden rule, which allows us to ascertain energy, spatial and angular distribution of hot carriers generated by different absorption mechanism. It allows us to make estimates of the efficiency of injection of the photo-generated carriers and to suggest the ways toward its improvement.

## Background: Drude theory of free carrier absorption

Currently in order to estimate the FCA in metal one uses the standard expressions for the AC conductivity using Drude formulae for the dielectric constant of metal

$$\varepsilon(\omega) = \varepsilon_b - \frac{Ne^2}{\varepsilon_0 m(\omega^2 + j\omega\gamma)} \tag{2}$$

where $\varepsilon_\infty$ is the relative permittivity due to bound carriers (inter-band transitions), m is the effective mass and the scattering rate $\gamma$ is usually introduced phenomenologically. The power dissipation rate is then

$$\frac{dU}{dt} = -\omega\varepsilon_0 \, \text{Im}[\varepsilon(\omega)]\frac{E^2}{2} = \frac{Ne^2\gamma}{m\omega^2}\frac{E^2}{2} = -\varepsilon_0 \frac{\omega_p^2}{\omega^2}\gamma\frac{E^2}{2} \tag{3}$$

Note here – the total energy density in the metal is

$$U = \frac{1}{4}\varepsilon_0 \, \text{Re}\left(\omega\frac{\partial\varepsilon}{\partial\omega}\right) + \frac{1}{4}\mu_0 H^2 = \frac{1}{4}\varepsilon_0\varepsilon_\infty E^2 + \frac{1}{4}\mu_0 H^2 + \frac{1}{4}\varepsilon_0 \frac{\omega_p^2}{\omega^2} E^2 \tag{4}$$

where the last term is the density of kinetic energy of carriers, so the power dissipation is

$$\frac{dU}{dt} = -2\gamma U_K \tag{5}$$

which of course makes a perfect sense since the energy gets lost only when the carriers are in motion. The scattering rate is usually written as a sum of different scattering processes, namely scattering on phonons, on various lattice imperfections, and due to electron-electron interaction. However, it remains questionable how one can apply theory developed for low frequencies, typically for the radio-frequencies all the way to visible light and even UV. At low frequencies all the processes take place in the vicinity of Fermi surface, so one only need to know the material properties near the Femi surface, while at higher frequencies almost all of the Brillouin zone starts playing role. This fact leads to interesting observations –some materials that are good low loss conductors in the RR region, such as Cu may suffer from higher losses in optical region than say Al, which is not such a great conductor. Clearly, scattering rate $\gamma$ is a strong function of frequency and in this work we establish this dependence and with it we re-derive Drude formula for the frequency dependence of dielectric constant.

## Is there a difference between absorption of photon and plasmon?

Before proceeding it is important to establish once and for all what is the difference between the hot carrier generation as a result of photon or plasmon decay. To understand it, it is important to realize that what is normally referred to as a "plasmon" is in reality a (surface) plasmon polariton, or SPP, i.e. a coupled mode (or a quasiparticle) in which energy is split between the energy of electro-magnetic field $\frac{1}{4}\varepsilon_0 E^2 + \frac{1}{4}\mu_0 H^2$ and the matter. In the metal, at frequencies significantly less than the plasma frequency, the energy of the matter is mostly the kinetic energy of the collective motion of free carriers $\frac{1}{4}\varepsilon_0 \left( \omega_p^2 / \omega^2 \right) E^2$, and also a contribution from the potential energy of the collective motion of bound carriers $\frac{1}{4}\varepsilon_0 (\varepsilon_\infty - 1) E^2$. In the dielectrics what is normally referred as a "photon" is in fact also a polariton and the energy is split between the energy of the electro-magnetic field and the energy of the bound carriers. The only difference between the plasmon and photon in dielectric is the fact that in plasmon a significant part of the energy is contained in the form of kinetic energy of free carriers while in the photon in dielectric energy is contained in the potential energy of bound carriers and in the magnetic field [19]. The distribution of energies is shown in Fig.2. The interaction between the polariton and the single electron-hole pairs excitations in metal or dielectric is still characterized by the same Hamiltonian $e\bm{r} \cdot \bm{E}$ (or $e\bm{p} \cdot \bm{A} / m$ in the momentum gauge) i.e. only the electric field of polariton interact with the matter.

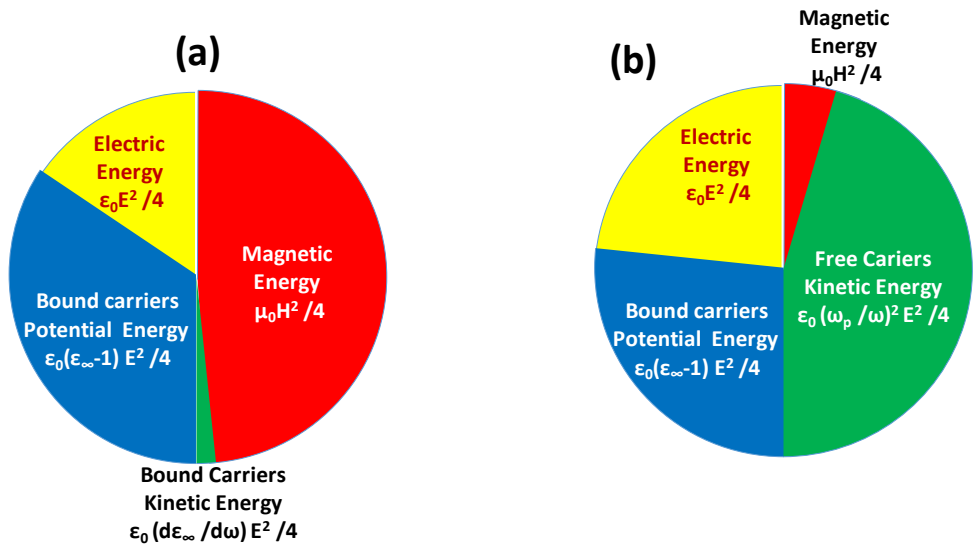

Fig.2 Energy breakdown in the (a) photon polariton in dielectric and (b) plasmon polariton in metal. The only difference is in relative contributions of magnetic and kinetic energies, and only the electric field determines the interaction between polariton and the single particle excitations in the matter.

When it comes to comparison between absorption of localized SPPs and the incoming photons in metal once again there is no physical difference between two processes – in fact if localized SPP can be considered a bound state the photon propagating in the vicinity of metal can be considered as simply an un-bound state. Of course, the actual loss will depend on the electric field distribution in the states, and the absorption will differ as a result, but the difference is quantitative and not qualitative. However, different absorption mechanisms will dominate in the bulk of the metal and near its surface and their relative strength will affect the properties of phot generated hot carriers, whether they are generated by SPPs or by incoming photons penetrating the metal. In the subsequent sections we consider these mechanisms one by one.

## Free carrier absorption arising from phonon and defect scattering

This is the most ubiquitous mechanism responsible for the intra-band free carrier absorption, and can be understood phenomenologically as simple friction accompanying the motion of free carries in the electro-magnetic field. For this reason, it is often referred to as "classical"[18] or "resistive"[16] and somehow not leading to hot carrier generation. This is strictly speaking not correct as of course hot carriers do get generated in this process that can be perfectly well described using quantum mechanical formalism as shown below.

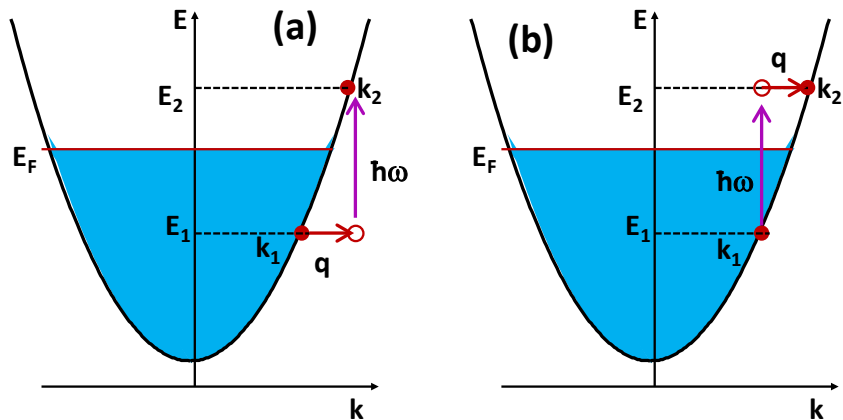

Fig.3 Two processes involved in phonon-assisted absorption in metal: (a) Phonon scattering in to the virtual state is followed by photon (SPP) absorption into the real state $k_2$ (b) Photon (SPP) absorption into virtual state is followed by scattering into the real state $k_2$

Consider a simple parabolic conduction band of a metal shown in Fig.3. When the photon (SPP) with energy $\hbar\omega \gg k_B T$ is inside the metal momentum conservation rule does not allow a direct absorption of it. The momentum mismatch can be provided by either phonon or impurity (imperfection). Here we consider a phonon with energy $\hbar\omega_p \ll \hbar\omega$ and wavevector $\boldsymbol{q}$ that provides momentum matching between the initial state of electron with energy and momentum $(E_1, \boldsymbol{k}_1)$ and the final state $(E_2 = E_1 + \hbar\omega, \boldsymbol{k}_2 = \boldsymbol{k}_1 \pm \boldsymbol{q})$. The $\pm$ sign corresponds to phonon emission and absorption respectively. Imperfections can be treated essentially as superposition of phonons with vanishingly low frequency and different wavevectrors. The interaction between the electromagnetic field and the electron is described by the Hamiltonian $H_{ef} = e\boldsymbol{A} \cdot \boldsymbol{v}$ where $\boldsymbol{A} = \boldsymbol{E}/\omega$ is the vector potential and $\boldsymbol{v} = \hbar \boldsymbol{k}/m$ (assuming parabolic band) is the electron velocity. The electron-phonon interaction is described by the Hamiltonian

$$H_q = D\sqrt{\frac{2\hbar\omega_p(n_p + 1/2 \pm 1/2)}{\rho V v_s^2}} \tag{6}$$

where $D$ is a deformation potential, $\rho$ is the density, $v_s$ is the sound velocity, and $n_p(T)$ is the temperature-dependent number of phonons with a given frequency. The phonons involved typically have wavevector q on the order of Fermi wavevector hence their frequency is on the order of Debye frequency of a few THz and we can introduce average parameter $\langle n_p \rangle$ which describes both absorption and emission.

What is important to consider is that the process of phonon assisted absorption can occur in two alternative sequences, or pathways. In the first pathway shown in Fig.3.a at first electron scatters from $\boldsymbol{k}_1$ to a virtual state $(E_1, \boldsymbol{k}_2)$ with the help of phonon and the energy is not conserved, with the deficit of energy $(-\hbar\omega)$. Following that photon gets absorbed and the electron goes into the final state $(E_2, \boldsymbol{k}_2)$. In the second pathway shown in Fig.3.b the electron first absorbs the photon and ends up in the virtual state $(E_2, \boldsymbol{k}_1)$ which now has excess energy $(-\hbar\omega)$. This is followed by a phonon assisted transition into the final state $(E_2, \boldsymbol{k}_2)$. The amplitudes of both pathways will interfere, hence, according to Fermi Golden rule, one can write for the absorption rate of photon by state $(E_1, \boldsymbol{k}_1)$ with spin conservation

$$\begin{aligned} R_{k_1} &= \frac{2\pi}{\hbar} \sum_{k_2} \left[ \frac{1}{2} \frac{(H_q)(e\boldsymbol{A} \cdot \boldsymbol{v}_2)}{\hbar\omega} - \frac{1}{2} \frac{(e\boldsymbol{A} \cdot \boldsymbol{v}_1)(H_q)}{\hbar\omega} \right]^2 \delta(E_2 - E_1 - \hbar\omega) = \\ &= \frac{2\pi e^2}{\hbar^3 \omega^4} V \frac{1}{8\pi^3} \int H_q^2 \left[ \frac{\boldsymbol{E}}{2} \cdot (\boldsymbol{v}_2 - \boldsymbol{v}_1) \right]^2 \delta(E_2 - E_1 - \hbar\omega) d\boldsymbol{k}_2 \end{aligned} \tag{7}$$

where we have gone from the summation to integration in k –space.

It is important that the change of the velocity of electron, i.e. its acceleration vector is parallel to the electric field as would follow from the classical theory. Next, when we perform integration in polar coordinates we can perform angle averaging of both phonon scattering strength and the direction and magnitude of $v_2 - v_1$

$$\left\langle \left( v_{2,z} - v_{1,z} \right)^2 \right\rangle = v_F^2 \left\langle \left( \cos\theta_2 - \cos\theta_1 \right)^2 \right\rangle_{\theta_1,\theta_2} \approx v_F^2 \times \left( \left\langle \cos^2\theta_1 \right\rangle_{\theta_1} + \left\langle \cos^2\theta_2 \right\rangle_{\theta_2} \right) = \frac{2}{3} v_F^2 \tag{8}$$

where $v_F$ is a Fermi velocity and obtain

$$R_{k1} = \frac{2\pi e^2}{\hbar^3 \omega^4} V \frac{1}{8\pi^3} \frac{E^2}{6} v_F^2 \int H_q^2 \delta(E_2 - E_1 - \hbar\omega) dk_2 \approx \frac{\pi}{3} \frac{e^2 E^2 v_F^2}{\hbar^3 \omega^4} V H_q^2 \rho_{E_1+\hbar\omega} \tag{9}$$

where $\rho_{E_1+\hbar\omega} = (1/8\pi^3) \int \delta(E_2 - E_1 - \hbar\omega) dk_2$ is the density of final states. We can now introduce the average electron-phonon scattering time as

$$\tau_{ep}^{-1}(\omega) = \left\langle (2\pi/\hbar) H_q^2 V \rho_{E_1+\hbar\omega} \right\rangle_{E_1} \tag{10}$$

and then obtain for the rate of absorption from the given energy level $E_1$

$$R_{E_1} = \frac{e^2 v_F^2 \tau_{ep}^{-1} E^2}{6\hbar^2 \omega^4} \tag{11}$$

Next we calculate the rate of hot carrier generation from all levels, i.e.

$$\frac{dN_e}{dt} = \int_{E_F-\hbar\omega}^{E_F} R_{E_1} \rho(E_1) dE_1 \approx \hbar\omega R_{E_1} \rho(E_F) = \frac{e^2 v_F^2 \tau_{ep}^{-1}(\omega) E^2}{6\hbar \, \omega^3} \rho(E_F) \tag{12}$$

Then the energy relaxation rate is

$$\frac{dU_\omega}{dt} = -\hbar\omega \frac{dN_e}{dt} \approx \frac{e^2 v_F^2 \tau_{ep}^{-1}}{6\omega^2} \rho(E_F) E^2 = \frac{e^2 v_F^2}{3\omega^2} \rho(E_F) \tau_{ep}^{-1} \frac{E^2}{2} \tag{13}$$

To make (13) look like (3) we multiply and divide by the square of plasma frequency

$$\frac{dU_\omega}{dt} = -\frac{e^2 v_F^2}{3\varepsilon_0 \omega_p^2} \rho(E_F) \varepsilon_0 \frac{\omega_p^2}{\omega^2} \tau_{ep}^{-1} \frac{E^2}{2} = -\gamma_{ph} \varepsilon_0 \frac{\omega_p^2}{\omega^2} \frac{E^2}{2} \tag{14}$$

where

$$\gamma_{ph} = \frac{e^2 v_F^2}{3\varepsilon_0 \omega_p^2} \rho(E_F) \tau_{ep}^{-1} \qquad (15)$$

However, for the spherical band the AC conductivity can be found by integrating over the Fermi surface as

$$\sigma(\omega) = i\frac{1}{3}\frac{e^2 v_F^2 \rho(E_F)}{\omega} = i\varepsilon_0 \frac{\omega_p^2}{\omega} \qquad (16)$$

Therefore, plasma frequency is

$$\omega_P^2 = \frac{1}{3}\frac{e^2 v_F^2 \rho(E_F)}{\varepsilon_0} \qquad (17)$$

And (15) becomes simply

$$\gamma_{ph}(\omega) = \tau_{ep}^{-1}(\omega) \qquad (18)$$

which is of course an intuitive result, but it should be noted that this results is frequency dependent as the scattering is averaged over the range of energies from the Fermi level up to $E_F + \hbar\omega$ and thus can be somewhat different from the scattering rate at Fermi level that enters the expression for conductivity at low frequencies. One can make a rough estimate of (10), assuming deformation potential $D \sim 10 eV$, density of states $\rho_{E_F} \approx \frac{3}{2} N_e / E_F$, where electron density is $6 \times 10^{22} cm^{-3}$, mass density $\rho \sim 20 g/cm^3$, speed of sound $v_s = 3.2 \times 10^3 m/s$, and the mean phonon energy $\langle \hbar\omega_p \rangle \sim 10 meV$. equal to Debye energy. One obtains for gold at room temperature $\gamma_{ph} = 6 \times 10^{13} s^{-1}$ Recent calculations [20] estimate room temperature values of $\gamma_{ph} = 3 \times 10^{13} s^{-1}$ for silver and $\gamma_{ph} = 1 \times 10^{14} s^{-1}$ for gold.

Next, it is important to estimate the directionality of the photo-excited hot carriers. What we need is to ascertain the photo-excitation dependence on the angle $\theta_2$ while averaging over the angle $\theta_1$, or

$$R(\theta_2) \sim \frac{\langle (v_{2,z} - v_{1,z})^2 \rangle_{v_1}}{\langle (v_{2,z} - v_{1,z})^2 \rangle_{v_1,v_2}} \sim \frac{\langle (\cos\theta_2 - \cos\theta_1)^2 \rangle_{\theta_1}}{\langle (\cos\theta_2 - \cos\theta_1)^2 \rangle_{\theta_1,\theta_2}} = \frac{\cos^2\theta_2 + \frac{1}{3}}{\frac{2}{3}} = \frac{1}{2}(3\cos^2\theta_2 + 1) \qquad (19)$$

So, the distribution is not uniform and is skewed along the direction of the electric field, which, for many plasmonic structures is directed normal to the surface. We shall consider the impact of this directionality later on (Fig.7).

## Absorption assisted by electron-electron scattering

This is the process that is often overlooked in the studies of hot carrier generation in plasmonics. Partially it is due to the fact that it is a second order process involving two electrons. In the field of photonics, most closely related to plasmonics, the higher order processes involving multiple photons are indeed often disregarded since they are quite weak and manifest themselves only at high powers, as is known to anyone familiar with nonlinear optics [21]. The situation, however, could not be more different in the condensed matter physics where multi-phonon and multi-electron interactions play the key role in many important phenomena, such as, for instance, heat transport[22]. The difference between optics and condensed matter physics is of course in the density of states, that is at least 10 orders of magnitude higher in the solid state than in optics. The large number of states available for scattering makes multi-particle processes important, and electron-electron scattering-assisted absorption of photons is just one of these processes. At low frequencies these processes are not important since the phase space available for scattering is within a narrow layer $\hbar\omega$ from the Fermi surface, but at higher photon frequencies this phase space increases proportionally to $\hbar\omega$, and, since there are two electrons involved in scattering the probability of this process increases as $\omega^2$. This fact has been observed in multiple experiments dating back to 1970's and 1980's [23-25], and theoretical calculations [26] performed on the basis Landau's Fermi liquid theory [27] have shown that indeed the electron-electron scattering rate at high frequencies can be evaluated as

$$\gamma_{ee,\omega}(\omega,T) \sim \gamma_{ee,0}(T)(\hbar\omega/2\pi k_B T)^2 \tag{20}$$

where for gold $\gamma_{ee,0} \sim 2\times 10^{11} s^{-1}$ at room temperature [23]. For the energy of 1eV the rate of electron-electron scattering involved in photon absorption $3\times 10^{13} s^{-1}$, while calculations performed more recently [20] predict somewhat higher rates of $7\times 10^{13} s^{-1}$ which is commensurate with phonon-assisted absorption $\gamma_{ph}$.

A rigorous theoretical description of the electron-electron scattering process is quite involved, and should follow full treatment of Fermi liquid [28], so in this section we outlined the physics of the process and perform a simplified and physically transparent order-of magnitude estimation of absorption aided by the electron-electron scattering.

There are four different pathways, all shown in Fig.4 that correspond to absorption of photon $\hbar\omega$ and simultaneous transition of two electrons (of different spins) from initial state $(E_1, \boldsymbol{k}_1; E_3, \boldsymbol{k}_3)$ to the final state $(E_2, \boldsymbol{k}_2; E_4, \boldsymbol{k}_4)$. Since this is elastic scattering the energy is conserved, $E_1 + E_3 + \hbar\omega = E_2 + E_4$. The momentum is also conserved momentum $\boldsymbol{k}_1 + \boldsymbol{k}_3 - \boldsymbol{k}_4 - \boldsymbol{k}_2 = K\boldsymbol{G}$, where $\boldsymbol{G}$ is one of the inverse lattice basis wavevector. Two electrons get scattered by the screened Coulomb potential

$$H_{ee,K} = f_K \frac{e^2}{\varepsilon_0 V(\Delta k^2 + k_s^2)} \tag{21}$$

where $k_s$ is the screening wavevector that for most metals is close to the Fermi wavevector $k_F$ and $\Delta k = k_3 - k_1$ is the momentum transfer between the two electrons. Since $\Delta k < k_F \sim k_s$, in order to make order-of magnitude estimate one can take

$$H_{ee,K} \approx f_K \frac{e^2}{\varepsilon_0 V k_F^2} \qquad (22)$$

When K=0 the process is Normal (N), otherwise it is an Umklapp (U) process [22] and the matrix element of Hamiltonian in (21) is reduced by the overlap integral $f_K = \int u^2(r) e^{jKG \cdot r} dr$ where $u(r)$ is the periodic part of Bloch function and the integral is taken over one unit cell. Obviously $f_0 = 1$. Let us now consider four sequences as in Fig.4 showing that there are four different sequences for absorption involving two electrons of different spins.

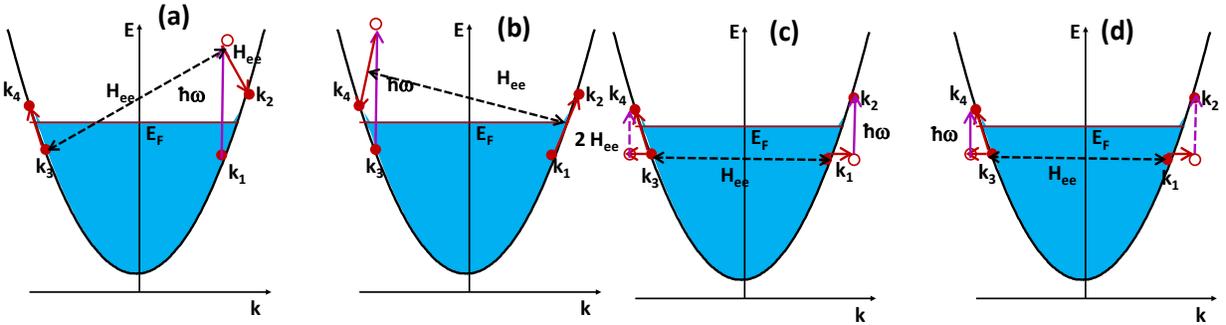

Fig.4 Absorption assisted by electron-electron scattering. Four different pathways resulting in generation of two electron hole pairs of different spins are shown.

(a) Photon excites the electron on the right transfers the system into the virtual state $(E_2, k_1; E_3, k_3)$ with an excess energy $\hbar\omega$ then electron-electron scattering moves it into the final state $(E_2, k_2; E_4, k_4)$

(b) Photon excites the electron on the left transferring the system into the virtual state $(E_1, k_1; E_4, k_3)$ with an excess energy $\hbar\omega$ then electron-electron scattering moves it into the final state $(E_2, k_2; E_4, k_4)$

(c) Electron-electron scattering transfers the system into the virtual state $(E_1, k_2; E_3, k_4)$ then photon interact with the electron on the right and moves the system into its final state $(E_2, k_2; E_4, k_4)$

(d) Electron-electron scattering transfers the system into the virtual state $(E_1, k_2; E_3, k_4)$ then photon interact with the electron on the left and moves the system into its final state $(E_2, k_2; E_4, k_4)$

The interference of four pathways in Fig.4 results in the following probability for the absorption of phonon or exciton by the electron in state $\mathbf{k}_1$

$$R_{k_1} = \frac{2\pi}{\hbar}V^3 \sum_{k_3}\sum_{k_4}\sum_{k_2}\left[\frac{(e\mathbf{A}\cdot\mathbf{v}_1)(H_{ee,K})}{-\hbar\omega} + \frac{(e\mathbf{A}\cdot\mathbf{v}_3)(H_{ee,K})}{-\hbar\omega} + \frac{(H_{ee,K})(e\mathbf{A}\cdot\mathbf{v}_2)}{\hbar\omega} + \frac{(H_{ee,k})(e\mathbf{A}\cdot\mathbf{v}_4)}{\hbar\omega}\right]^2 \times$$
$$\times \delta(\mathbf{k}_3+\mathbf{k}_1-\mathbf{k}_4-\mathbf{k}_2-K\mathbf{G})\delta(E_1+E_3+\hbar\omega-E_2-E_4) =$$
$$\frac{2\pi e^2}{\hbar^3\omega^4}V^3\sum_{k_3}\sum_{k_4}\sum_{k_2}H_{ee,K}^2\left[\mathbf{E}\cdot(\mathbf{v}_3+\mathbf{v}_1-\mathbf{v}_4-\mathbf{v}_2)\right]^2\delta(\mathbf{k}_3+\mathbf{k}_1-\mathbf{k}_4-\mathbf{k}_2-K\mathbf{G})\delta(E_1+E_3+\hbar\omega-E_2-E_4)$$

(23)

We now see that for the parabolic band N-process with K=0 results in no absorption as the sum of velocities is equal to zero. That is precisely the classical result that follows from the momentum conservation – the elastic electron scattering conserves the total momentum of electrons and since velocity is proportional to momentum the total velocity and current are conserved indicating zero resistance and hence zero energy dissipation. Therefore, one should only consider U processes, realistically only with the order K=1 so that one of the generated hot carrier ($k_2$) ends up in the adjacent Brillouin zone as shown in Fig.5a. Obviously the velocity is no longer conserved and the energy gets dissipated. Of course, another possibility is that the second electron $k_4$ ends up in the adjacent Brillouin zone.

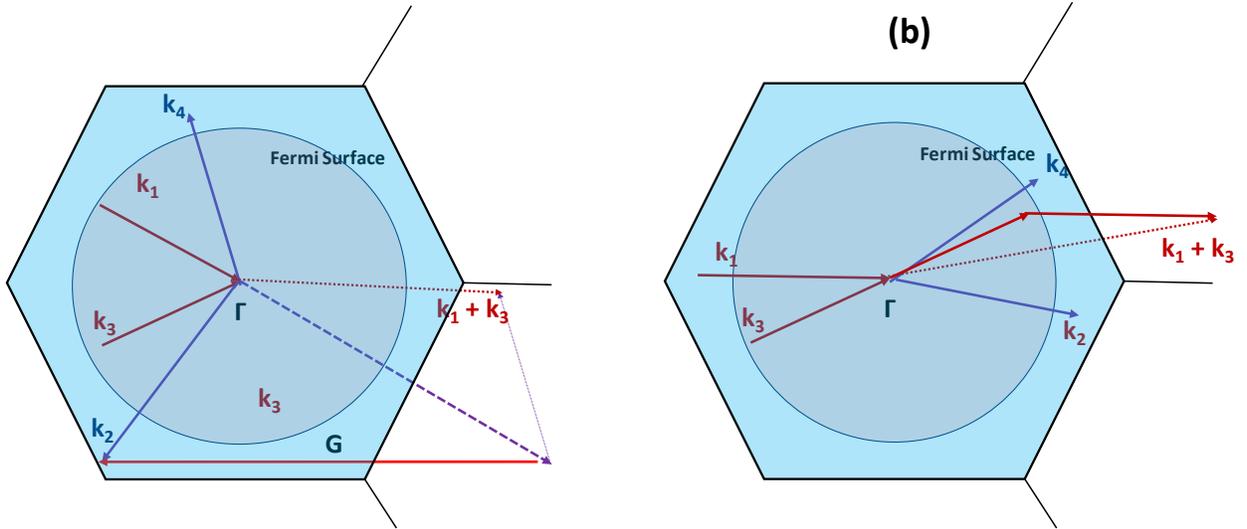

Fig.5 (a) Photon (SPP) absorption with energy $\hbar\omega$ assisted by Umklapp scattering of electrons. As photon (SPP) gets annihilated the electrons l$\mathbf{k}_1,\mathbf{k}_3$ below the Fermi surface get scattered into the states $\mathbf{k}_2,\mathbf{k}_4$ above Fermi surface with momentum conservation preserved by the lattice vector $\mathbf{G}$ . (b) Thermalization of a hot electron. . A hot electron $\mathbf{k}_1$ with energy $\hbar\omega$ above Fermi level and a cold electron $\mathbf{k}_3$ scatter into the states $\mathbf{k}_2,\mathbf{k}_4$ above Fermi surface each with energy less than $\hbar\omega$. Both normal and Umklapp processes are possible. The similarity between two processes indicate that the rate e-e scattering assisted absorption is commensurate with the rate of hot electrons thermalization.

Therefore, the rate of absorption from a particular state k₁ is obtained by integration over all the possible initial and final states of the second electron, while the state k4 is determined by the momentum conservation.

$$R_{k_1} = \frac{2\pi e^6 E^2 (\hat{e}\cdot G)^2}{k_F^4 \varepsilon_0^2 \hbar m^2 \omega^4} \left(\frac{1}{4\pi^3}\right)^2 \int_{k_3}\int_{k_2} \left[A_U(k_2)f_1^2(k_2) + A_U(k_4)f_1^2(k_4)\right]\delta(E_1+E_3+\hbar\omega-E_2-E_4)dk_2 dk_3 \tag{24}$$

where $\hat{e}$ is a unit vector denoting polarization. Of course, integration assumes that the energies of the initial states 1 and 3 are below Fermi surface while the energies of final states 2 and 4 are above it. Let us now elaborate on the nature of the coefficients $A_U(k)$ and $f_1^2(k)$. The first coefficient $A_U(k_2)(A_U(k_4)=1$ when the final wavevector $k_2(k_4)$ is outside the first Brillouin zone and is equal to zero otherwise, while the overlap integral

$$f_1(k) = \int u_{k-g}^*(r)e^{iG\cdot r}u_k(r)dr \tag{25}$$

is essentially the first order Fourier component of the periodic Bloch function $u_k(r)$ which for nearly free electrons in the noble metals obviously becomes large only in the vicinity of the Brillouin zone boundary, hence the whole term in the square bracket is substantial when one of the final wavevectors exceeds the Brillouin zone boundary by a relatively small value.

Next, following example (10) we shall introduce the effective electron-electron Umklapp scattering rate as

$$\tau_{ee,U}^{-1}(\omega) = \frac{2\pi e^4}{\hbar \varepsilon_0^2 k_F^4}\left(\frac{1}{4\pi^3}\right)^2 \int_{k_3}\int_{k_2}\left[A_U(k_2)f_1^2(k_2)+A_U(k_4)f_1^2(k_4)\right]\delta(E_1+E_3+\hbar\omega-E_2-E_4)dk_2 dk_3 \approx$$

$$\approx F_U(\hbar\omega) \times \tau_{ee}^{-1}(\omega) \tag{26}$$

where

$$\tau_{ee}^{-1}(\omega) = \frac{2\pi e^4}{\hbar\varepsilon_0^2 k_F^4}\left(\frac{1}{4\pi^3}\right)^2 \int_{k_3}\int_{k_2}\delta([E_F+\hbar\omega]+E_3-E_2-E_4)dk_2 dk_3 \tag{27}$$

is the electron-electron relaxation (thermalization) rate of the hot carriers having energy of $\hbar\omega$ above the Fermi (as shown in Fig. 5b) level for which a large amount of experimental data has been accumulated [29-33], and

$$F_U(\hbar\omega) = \left\langle A_U(k_2)f_1^2(k_2) + A_U(k_4)f_1^2(k_4)\right\rangle_{k_2,k_4} \tag{28}$$

is the relative strength of Umklapp process. It is important to note that just in addition to the scattering of two different electrons with different spins there is also a contribution from the scattering of two identical electrons which is smaller due to exchange. This interaction has been considered in [34] and found to be same order but somewhat less (order of 1/2) than interaction of carriers with different spins. As long as exchange interactions are factored into the expression (27) for the electron thermalization they will also be factored in the expression for the photon (plasmon) decay obtained from (24)

$$R_{E_1} = \frac{e^2 G^2}{3m^2 \omega^4} F_U(\hbar\omega)\tau_{ee}^{-1}(\hbar\omega)E^2 \tag{29}$$

where the factor (1/3) has been obtained by averaging over the directions of the reciprocal lattice $G$. Next following all the steps taken for the phonon-assisted absorption in (12) -(14) we obtain

$$\frac{dU_\omega}{dt} = -\hbar^2 \omega^2 R_{E_1} \rho(E_F) \approx \frac{\hbar^2 e^2 G^2}{3m^2 \omega^2} \rho(E_F) F_U(\hbar\omega)\tau_{ee}^{-1}(\hbar\omega)E^2 \tag{30}$$

Then using (17) we transform this expression into

$$\frac{dU_\omega}{dt} = 2\frac{\hbar^2 G^2}{m^2 v_F^2} F_U(\hbar\omega)\tau_{ee}^{-1}(\hbar\omega)\varepsilon_0 \frac{\omega_P^2}{\omega^2}\frac{E^2}{2} \tag{31}$$

Now the term $2\hbar^2 G^2 / m^2 v_F^2$ is commensurate with the ratio of the energy at the boundary of the BZ to the Fermi energy and is on the order of unity for the metal, hence it can also be incorporated into the "Umklapp factor" $F_U(\hbar\omega)$ and according to (3) one finally obtains the expression for the effective electron-electron scattering that can be entered into the Drude formula as

$$\gamma_{ee}(\omega) = F_U(\hbar\omega)\tau_{ee}^{-1}(\hbar\omega) \tag{32}$$

Now, the term $\tau_{ee}^{-1}(\hbar\omega)$ has been studied in numerous theoretical [35-38] and experimental [29-33] works which all show that for the energies of 1eV above Fermi level it is on the scale of $10^{14}$ s$^{-1}$ and here we should ascertain whether this order of magnitude follows from our theoretical model

Following Ziman's derivation of the AC resistivity in [39] we obtain the expression for the integral in (27) as roughly

$$I = \int_{k_3}\int_{k_2} \delta([E_F + \hbar\omega] + E_3 - E_2 - E_4)d\mathbf{k}_2 d\mathbf{k}_3 \approx \frac{\pi^4 k_F^6}{192}\frac{1}{E_F}\left(\frac{\hbar\omega}{E_F}\right)^2 \tag{33}$$

and then substitute it into (27) to obtain

$$\tau_{ee}^{-1}(\omega) = \frac{\pi k_F^2}{96\hbar}\left(\frac{e^2}{4\pi\varepsilon_0}\right)^2 \frac{1}{E_F}\left(\frac{\hbar\omega}{E_F}\right)^2 \tag{34}$$

This expression can be simplified by introducing Bohr radius

$$a_0 = \frac{4\pi\varepsilon_0 \hbar^2}{e^2 m_0} = 0.53 A \tag{35}$$

to obtain assuming $k_F \sim a_0^{-1}$

$$\tau_{ee}^{-1}(\omega) = \frac{\pi k_F^2}{96\hbar}\left(\frac{\hbar^2 k_F}{m_0 a_0}\right)^2 \frac{1}{E_F}\left(\frac{\hbar\omega}{E_F}\right)^2 \approx \frac{\pi}{24}\frac{E_F}{\hbar}\left(\frac{\hbar\omega}{E_F}\right)^2 \tag{36}$$

Using relation between plasma frequency and the Fermi energy, $\omega_P \approx \frac{4}{\sqrt{3\pi}}\frac{E_F}{\hbar}$ one finally obtains

$$\tau_{ee}^{-1}(\omega) \approx \frac{\pi^{3/2}\sqrt{3}}{96}\omega_p\left(\frac{\hbar\omega}{E_F}\right)^2 \tag{37}$$

This expression is close to one given by Pines in [28] where the numerical factor equals $\pi^2\sqrt{3}/128$. For gold one obtains $\tau_{ee}^{-1}$ ranging from $4.5\times 10^{13} s^{-1}$ at 1eV above Fermi level to $1.8\times 10^{14} s^{-1}$ at 2eV above it. The Umklapp factor $F_U(\hbar\omega)$ is not as simple to ascertain, but various calculations [34, 39] estimate it to be on the scale of 30 to 40% from which it follows that Umklapp electron-electron scattering rate $\gamma_{ee}(\omega)$ reaches $10^{13} s^{-1}$ at photon (SPP) energies less than 1eV and becomes dominate factor in absorption at energies exceeding 2eV, in agreement with experimental data [23]. Therefore, neglecting this process is erroneous. Since two electron-hole pairs are excited by absorption of a single photon (SPP) the average hot carrier energy is only $\hbar\omega/4$

## Absorption via Landau damping (a.k.a. surface collision assisted absorption)

Another way that momentum conservation in the transition between two states with different wave vectors can be restored is simply by using the "recoil" occurring when the electron gets reflected from the surface and the momentum is transferred between the electron and the wall (i.e. entire metal body). Phenomenological theory of this process has been first considered by Kreibig and Volmer [40], where they simply introduced a "surface collision scattering rate" $\gamma_s \sim v_F/a$ where $v_F$ is Fermi velocity a is the characteristic dimension of the metallic object. According to this phenomenological treatment the additional damping is caused by the limited physical dimension of the system and is the result of restriction of mean free path of electrons. Later it was shown that this process is related to Landau damping of electron [19, 41-44]. Landau damping[45] is the process of energy transfer between free electrons and electromagnetic waves occurring when their velocities are matched, for which the electro-magnetic wave must have larger (by a factor of 100 or so) wavevector than free space wavevector. While no propagating SPP can have such a large wavevector along its direction of propagation, lateral confinement of field means that large wavevectors are present in the spatial Fourier spectrum of the field. Here we estimate the rate of hot carrier generation via Landau damping and their properties, namely the energy, directionality and spatial positions, with some not entirely expected results.

Matrix element of Hamiltonian for the transition between two states separated by energy $\hbar\omega$ can be estimated using either the A.p or E.r Hamiltonian as

$$H_{12}^{E \cdot r} = -\frac{e}{j2\Delta kL}\int E(x)e^{+j\Delta kx}dx = -\frac{e}{j2\Delta kL}\tilde{E}(\Delta k) \tag{38}$$

which is essentially a Fourier transform of the electric field shape. Here (since the wavevectors in y and z directions are the same) $\Delta k = \omega/v_{Fx}$ and L is the quantization length. Next let us evaluate the Fourier transform. First, let us assume that the surface is really sharp and we can represent the electric field, shown in Fig. 6a as a product

$$E(x) = H(x)F(x) \tag{39}$$

of a step function $H(x)$ is whose Fourier transform shown in Fig. 6b is

$$\tilde{H}(k) = \pi\delta(k) - \frac{j}{k} \tag{40}$$

and the shape function $F(x)$ shown in Fig 5a by dashed line whose spatial extent is significantly broader than $1/\Delta k \sim v_F \cos\theta/\omega < 1nm$. Therefore its spatial spectrum $\tilde{F}(k)$ also shown in Fig 5b is much narrower than $\Delta k$

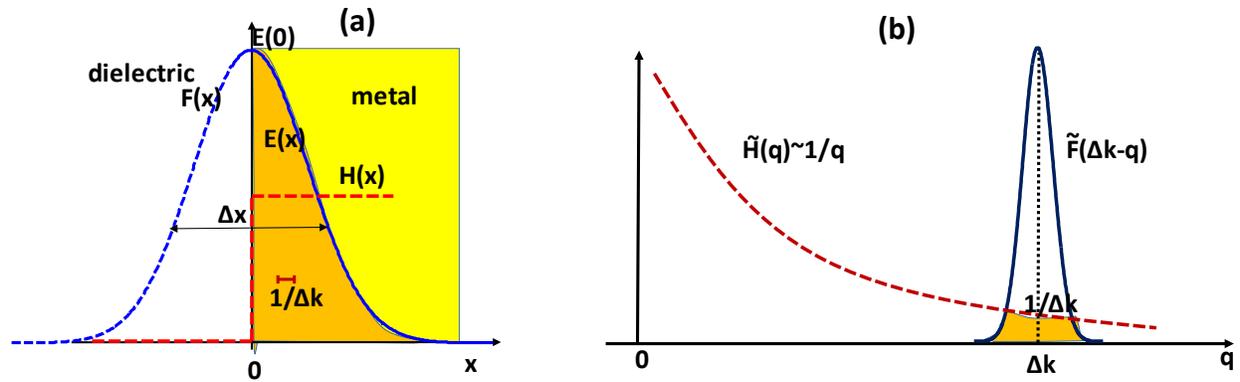

Fig.6 (a) Electric field E(x) (shaded) at the sharp interface between metal and dielectric (b) Fourier transform of the field in is a convolution with a step function

The Fourier transform of the field can then be found as a convolution of two Fourier transforms,

$$\tilde{E}(\Delta k) = \frac{1}{2\pi} \int_{-\infty}^{\infty} \tilde{H}(q)\tilde{F}(\Delta k - q)dq \approx \frac{1}{2}\tilde{F}(\Delta k) - \frac{j}{2\pi\Delta k}\int_{-\infty}^{\infty}\tilde{F}(q)dq \approx \frac{j}{\Delta k}E(0) \qquad (41)$$

In other words, the result is independent of the exact shape of the mode and depends only on the value of the field at the surface. Clearly in the absence of a sharp feature in the electric field profile the value of $\tilde{E}(\Delta k)$ would be far less and the Landau damping would become insignificant. And so we have

$$|H_{12}|^2 = \frac{e^2 E^2(0)}{4\Delta k^4 L^2} = \frac{e^2 E^2(0) v_F^4}{4\omega^4 L^2}\cos^4\theta \qquad (42)$$

As the next step we calculate the absorption rate from a given state with the wavevector $k_x$

$$R_{k1} = (2\pi/\hbar) H_{12}^2 V \rho_{E_1 + \hbar\omega} \qquad (43)$$

where $\rho_x(E_F) = (2\pi\hbar v_F |\cos\theta|)^{-1} L$ is one-dimensional density of the final states, evaluated under consideration that neither spin nor direction of propagation change as the transition takes place. For the change of direction the required wavevector would be $2k_F \gg \Delta k$ and therefore this process can be neglected. Thus we obtain

$$R_{k1} = \frac{1}{\hbar^2}\frac{e^2 E^2(0) v_F^3}{4\omega^4 L}|\cos^3\theta| \qquad (44)$$

Averaging over the directions $\langle|\cos^3\theta|\rangle_\theta = 1/4$ we obtain

$$R_{E1} = \frac{e^2 E^2(0) v_F^3}{4\hbar^2 \omega^4 L} \qquad (45)$$

The total rate of absorption than requires integration over all the density states within $\hbar\omega$ from Fermi level and volume. Integration over volume can be split over the integration of the quantization length L (which may be different for different transverse co-ordinates y and z, and then surface integration over y and z . What one obtains as a result is the total power transferred from the plasmons (electric field) to the single electron-hole excitations

$$\int \frac{dU}{dt}dV = -\frac{e^2 v_F^3}{16\omega^2}\rho(E_F)\int E_\perp^2 dS \qquad (46)$$

Where the integral on the right hand side is taken over the surface and $E_\perp$ is the component of the electric field that is normal to the surface. Using (17) we obtain

$$\int \frac{dU}{dt}dV = -\frac{3}{16}\frac{\omega_P^2}{\omega^2} v_F \int \varepsilon_0 E_\perp^2 dS \qquad (47)$$

At the same time, integration of (3) over the volume of the metal yields formally

$$\frac{dU}{dt}dV = -\frac{1}{2}\frac{\omega_p^2}{\omega^2}\int \gamma(\mathbf{r})\varepsilon_0 E(\mathbf{r})^2 dV = -\frac{\gamma_{eff}}{2}\frac{\omega_p^2}{\omega^2}\int_{metal} \varepsilon_0 E(\mathbf{r})^2 dV \tag{48}$$

$$(49)$$

Therefore, we may introduce the effective Landau damping (or surface collision) rate as

$$\gamma_s = \frac{3}{8} v_F / d_{eff} \tag{50}$$

where the effective surface to volume ratio is

$$d_{eff}^{-1} = \frac{\int \varepsilon_0 E_\perp^2(\mathbf{r})dS}{\int_{metal} \varepsilon_0 E(\mathbf{r})^2 dV} \tag{51}$$

A similar expression had been previously obtained in [46, 47]. The whole Landau damping rate can then be interpreted as simply the inverse of the time that takes the average electron propagating along the direction of electric field to reach the surface. For the SPP propagating along the interface plane between the metal and dielectric one obtains previously reported result $\gamma_s = \frac{3}{4}qv_F$ where q is the decay constant inside the metal, while for the spherical nanoparticle of the diameter $a$ the result is $\gamma_s = \frac{3}{4}v_F/a$ which is very close to the very intuitive interpretation giving by Kreibig [40].

## Directionality of excited carriers

We can now compare the directions of propagation of the carriers excited by three different mechanisms. For the Landau damping it follows from (44) that normalized distribution of photogenerated hot carriers is $R(\theta_2) \sim |\cos^3 \theta_2|/\pi$. For the phonon and defect assisted absorption according to (19) the angular distribution of hot carriers is $R(\theta_2) \sim (3\cos^2 \theta_2 + 1)/8\pi$. For the electron-electron scattering assisted absorption the distribution of hot carriers is omnidirectional $R(\theta_2) \sim 1/4\pi$. These distributions are shown in Fig. 7a. The impact of this difference in directionality is shown in Fig.7b where the injection efficiency $I.E.(\Theta_a) = \int_0^{\Theta_a} R(\theta_2)d\theta_2$ is shown as a function of the acceptance angle $\Theta_a$ defined in (1). One can see that the injection efficiency of the carriers generated by Landau damping is four times higher than that of omnidirectional carriers engendered by e-e scattering and twice as high as that carriers generated with the assistance of phonons and imperfections. This factor of 4 is important as it may be one of the factors that exp

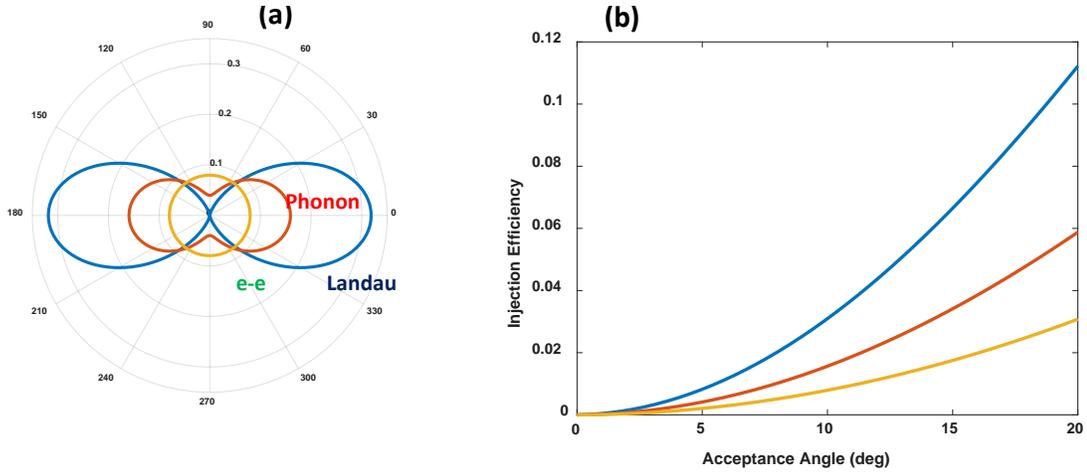

Fig.7 (a) angular distributions of the hot carriers photoexcited by different processes. (b) Injection efficiency of hot carriers

## Where do the hot carriers get excited?

Previous calculation has shown us that Landau damping originates from the momentum conservation violation at the abrupt boundary. It is also related to the phenomenon of non-locality [48-51], i.e. the dependence of the dielectric function of the free electron gas on the wavevector, or spatial dispersion that can be described by Lindhard's formula [52]

$$\varepsilon(\omega,k) = \varepsilon_b + \frac{3\omega_p^2}{k^2 v_F^2}\left[1 - \frac{\omega}{2kv_F}\ln\frac{\omega+kv_F}{\omega-kv_F}\right] \tag{52}$$

As one can see when $|k| \geq \omega/v_F \equiv \Delta k$ Landau damping becomes possible and one can write the expression for the imaginary part of the dielectric function as

$$\varepsilon_i(\omega,|k|>\Delta k) = 3\pi \omega_p^2 \omega / 2k^3 v_F^3 = \frac{3}{2}\pi \frac{\omega_p^2 \omega}{k^3 v_F^3}, \tag{53}$$

This spatial dispersion is shown in Fig. 8

For each value of the wavevector $k \geq \Delta k$ the spectral power density of the field according to (41)is

$$\left|\tilde{E}(\mathbf{k})\right|^2 \approx \frac{|E(0)|^2}{k^2} \tag{54}$$

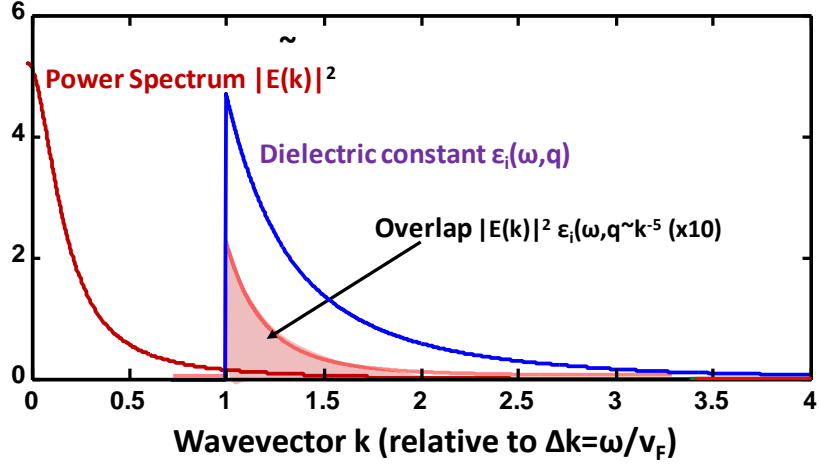

Fig.8 Calculating effective dielectric constant (imaginary part) and damping rate as an overlap of the spatial spectral power density of the field and the spatial spectrum of the imaginary part of dielectric constant

Only longitudinal component of the field $\tilde{E}_\parallel(\boldsymbol{k}) = \tilde{E}(\boldsymbol{k}) \cdot \boldsymbol{k}/k$ gets absorbed, hence, one can then evaluate the effective dielectric constant $\varepsilon_{eff,i}(\omega)$ as a function of the size and shape of an SPP mode by evaluating the overlap of with $\varepsilon_i(\omega,k)$ as seen in Fig.8

$$\varepsilon_{eff,i} = \frac{3\pi\omega_p^2 \omega}{2v_F^3} \int\limits_{k>\Delta k}^{\infty} k^{-3} \left|\tilde{E}_\parallel(k)\right|^2 d^3k \bigg/ \int\limits_{k>\Delta k}^{\infty} \left|\tilde{E}(k)\right|^2 d^3k \tag{55}$$

Splitting now integration first along the direction of electric field and then in plane plane normal and using Parseval theorem we obtain

$$\varepsilon_{eff,i} = \frac{3\pi\omega_p^2 \omega}{2v_F^3} \oint \left|E_\perp(0,\boldsymbol{r}_\perp)\right|^2 \int\limits_{|k|>\Delta k}^{\infty} k^{-5} dk dS \bigg/ 2\pi \int\limits_{metal} E(\boldsymbol{r})^2 dV = \frac{3}{8}\frac{\omega_p^2}{\omega^3} v_F \frac{\int E_\perp^2(\boldsymbol{r}) dS}{\int\limits_{metal} E(\boldsymbol{r})^2 dV} \tag{56}$$

and then it immediately follows from (2) that $\varepsilon_{eff,i} = \omega_p^2 \gamma_s / \omega^3$ and therefore $\gamma_s = \frac{3}{8} v_F / d_{eff}$, i.e. precisely the equation (50).

The analysis of Landau damping performed so far allows us to estimate the total rate of absorption, but by its nature the integral in (56) precludes us from determining where exactly the absorption takes place. A naïve way of thinking would lead us to conclusion that since the electron wavefunction is spread out along the entire metal volume, so does the absorption, which is absurd. In fact the absorption involves excitation of a wave packet near the interface, because that is where all the free electrons wavefunctions get locked in phase by the boundary conditions. Since the energies of photo-excited electrons are spread over $\hbar\omega$, their wavevectors are spread over the range $\Delta k = \omega/v_F$, hence one should expect that the absorption should be localized within roughly $\Delta L \sim 2\pi/\Delta k$ from the boundary.

To find the spatial distribution of the absorption we need to modify the expression for power dissipation (3) for the case of non-local medium whose dielectric constant is

$$\varepsilon(\mathbf{r}-\mathbf{r}_1) = \frac{1}{2\pi}\int \varepsilon(\mathbf{k})e^{j\mathbf{k}\cdot(\mathbf{r}-\mathbf{r}_1)}d\mathbf{k} \tag{57}$$

Power dissipation is then

$$\frac{dU(\mathbf{r})}{dt} = \frac{1}{2}\mathbf{E}(\mathbf{r})\cdot \mathbf{J}_r(\mathbf{r}) = \frac{1}{2}\omega \mathbf{E}(\mathbf{r})\cdot \mathbf{P}_i(\mathbf{r}) \tag{58}$$

where $\mathbf{J}_r$ is in-phase (real) part of current density and $\mathbf{P}_i = \omega^{-1}\mathbf{J}_i$ is the quadrature (imaginary) part of the polarization which can be expresses as

$$\mathbf{P}_i(\mathbf{r}) = \varepsilon_0 \int \varepsilon_i(\mathbf{r}-\mathbf{r}_1)\mathbf{E}(\mathbf{r}_1)d\mathbf{r}_1 \tag{59}$$

or, taking Fourier transform of this convolution integral

$$\tilde{\mathbf{P}}_i(\mathbf{k}) = 2\pi\varepsilon_0\varepsilon_i(\mathbf{k})\tilde{\mathbf{E}}_\|(\mathbf{k}) \tag{60}$$

Finally taking inverse Fourier transform of (60) and substituting it into (58) we obtain

$$\frac{dU(\mathbf{r})}{dt} = \frac{1}{2}\varepsilon_0\omega \mathbf{E}(\mathbf{r})\cdot \int \varepsilon_i(\mathbf{k})\tilde{\mathbf{E}}_\|(\mathbf{k})e^{j\mathbf{k}\cdot\mathbf{r}}d\mathbf{k} \tag{61}$$

Before proceeding with an example one should note that the expression for the dielectric constant (52) does not include the bulk losses $\gamma_b(\omega) = \gamma_{ph}(\omega) + \gamma_{ee}(\omega) + ...$ caused by scattering on phonons, defects, other electrons as well as residual inter-band absorption. A simple way to include them is to introduce the uncertainty of the wavevector commensurate with the inverse of the mean free path, i.e. $\delta k = L_{mfp}^{-1} = 4\gamma_b / 3v_F$. The reason for inclusion of factor 4/3 will be made clear below.

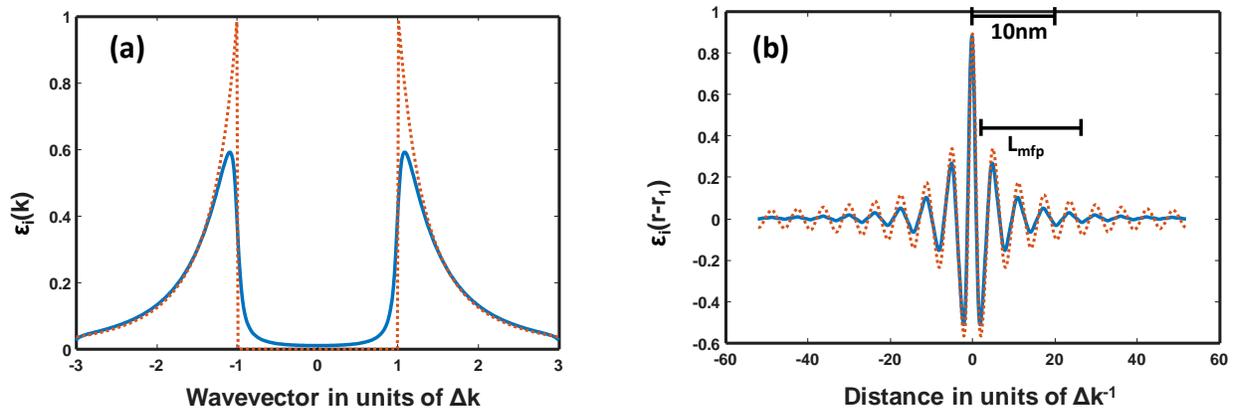

Fig.9 (a) Imaginary part of the metal dielectric constant according to Lindhard's formula (53) – dashed curve and modified for bulk scattering according to (62) –solid line. (b) Dielectric response of the metal in real space according to Lindhard's formula (53) – dashed curve and modified for bulk scattering according to (62) –solid line

A modified imaginary part of dielectric constant incorporating the bulk absorption is then a convolution

$$\varepsilon_i^{'}(\omega,k) = \int \frac{\pi^{-1}\delta k}{(k-k_1)^2 + \delta k^2} \varepsilon_i^{'}(\omega,k_1) dk_1 \qquad (62)$$

as shown in Fig.9a for the case of $\gamma_b = 10^{14} s^{-1}$ and $\omega = 2.5 \times 10^{15} s^{-1}$ ($\lambda = 754nm$) so that $\delta k / \Delta k = 4\gamma_b / 3\omega = 0.053$. As one can see, two obvious changes can be seen when bulk scattering is included. First of all, for small wavevectors $k << \Delta k$ one recovers precisely the Drude expression for the bulk dielectric constant (imaginary part) $\varepsilon_i^{'}(\omega,0) = \omega_p^2 \gamma_b / \omega^3$ (that is why a factor of 4/3 has been included in the uncertainty of momentum). Second, the sharp step of $\varepsilon(k = \pm\Delta k)$ has been smoothed by $\delta k$. From the properties of Fourier transform it follows now that in real space the dielectric response will decay over the length $\delta k^{-1}$, i.e. roughly one mean free path. The response in real space is shown in Fig.9b and it shows rapid oscillations with a period $\Delta L = 2\pi / \Delta k$ and decays nearly completely over $L_{mfp}$

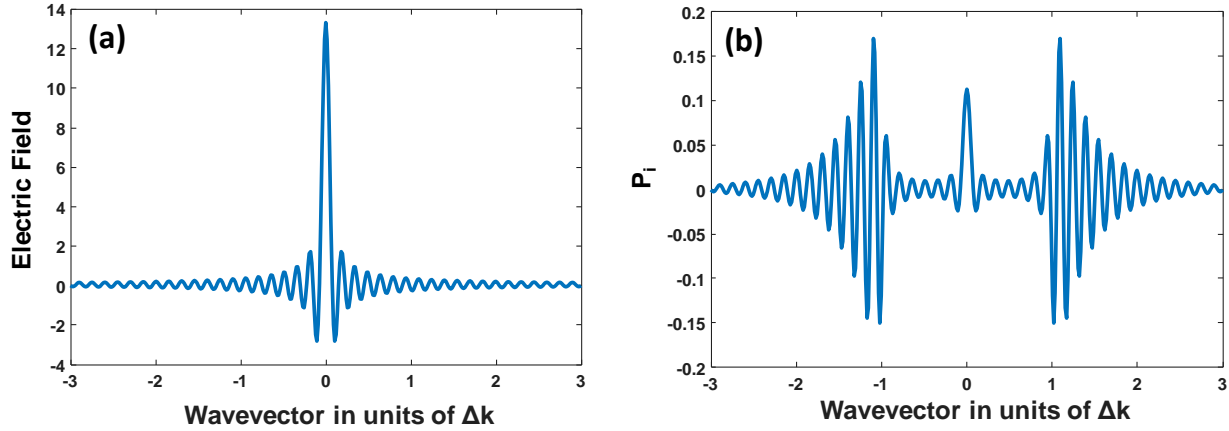

Fig.10 (a) Spatial spectrum of the electric field in the 40nm Au plate (b) Spatial spectrum of the imaginary part of polarization (in phase current)

Let us now consider a simple example of a thin metal plate of thickness d=40nm that is entirely penetrated by the electric field normal to its surface which results in the sinc-like spatial spectrum of it $\tilde{E}_{\parallel}(k)$ shown in Fig. 10a. and spectrum of the imaginary (quadrature) part of polarization (in phase current) (60) that is shown in Fig. 10b. The peak at k=0 represents bulk absorption and the oscillating wings at $|k| > \Delta k$ represent the effect of surface collisions (Landau damping).

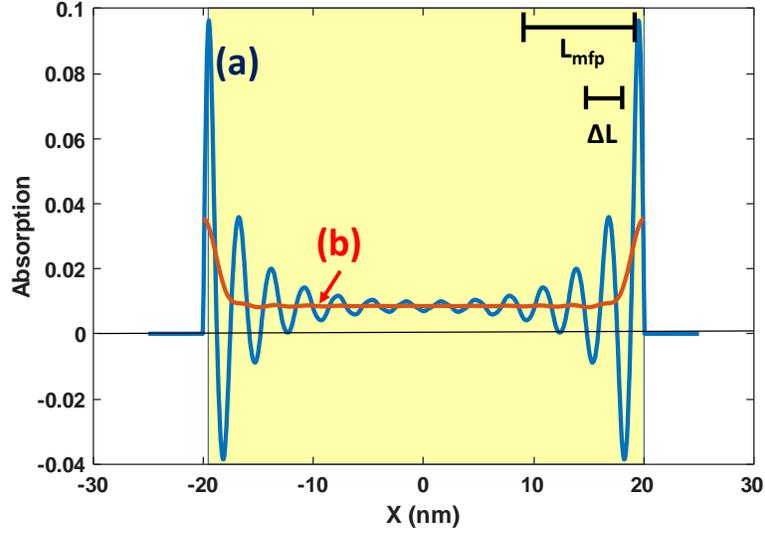

Fig.10 (a) Spatial dependence of energy dissipation rate in the 40nm thick Au plate. (b) Same but with oscillations smoothed out over each period.

Finally, in Fig.11 we show the rate of energy dissipation (61) as curve a it clearly exhibits strong oscillatory response with a period $\Delta L = 2\pi / \Delta k = 3nm$. These oscillations are due to singularity of the dielectric constant of metal at $k = \pm \Delta k$ and are conceptually similar to the Friedel oscillations of dielectric response at zero frequency. As expected, the oscillations decay away from the surfaces over one mean free path of about 12nm. In addition to oscillatory response one can see a constant response of bulk absorption. The physical meaning of oscillation simply means that electron-hole pairs created at the positive peaks move into the valleys where the absorption is then negative. But overall absorption is obviously positive. In fact, if we average the absorption over a period of oscillation (curve b) while keeping the integral under the curve constant, one can see that the Landau damping takes place entirely within $\Delta L = 2\pi / \Delta k = v_F T_{opt}$. This result simply means that all the carriers that can reach the surface within one optical period $T_{opt}$ can participate in the Landau damping. Also, one can easily observe from Fig. 11 that the ratio of Landau-damping caused free carrier absorption to the bulk absorption is about 0.3 which is of course simply the ratio of $\gamma_a / \gamma_b \sim L_{mfp} / d$.

## Final observations

In this work we have developed a simple quantum theory of free carrier absorption in metal and determined the rates of hot carrier generation, their energy, spatial position and angular distribution for different mechanisms. Here we can offer a short summary

Phonon and defect assisted absorption occurs throughout the entire volume of the SPP (or photon) mode in the metal. The rate of damping caused by this process is on the scale of $10^{13} - 10^{14} s^{-1}$. The energy of SPP is on average split evenly between the hole and electron thus each carrier on average has energy $\hbar\omega/2$. The hot electrons have their momentum preferentially directed along the electric field but the directionality is weak (factor of 2). Since only a fraction of hot electrons reaches the surface tis process is moderately useful for such applications as photo detection and photo catalysis.

Electron -Electron scattering assisted absorption should not be overlooked. This process is strongly dependent on the SPP (photon) energy and in the visible range its rate is comparable to that of phonon-assisted SPP damping. Since two pairs of carriers are generated, each carrier on average has energy of only $\hbar\omega/4$. Furthermore, the carriers are omnidirectional. Therefore, from the applications point of view this process is deleterious and may cause reduction of efficiency at shorter wavelengths

Interband absorption that onsets in visible for gold and in UV for silver has not been considered in this work, since it is obvious that this process does not generate useful carriers. Most of the SPP (photon) energy is consumed by promoting the electron from d-shell to the Fermi level – therefore the electron is generated not far above the Fermi level and typically does not have enough kinetic energy to overcome the barrier. The hole in d-shale has very low velocity and cannot reach the surface at all. For these reasons hot electron devices probably cannot operate with any degree of efficiency at short wavelengths.

Finally, Landau damping (or surface collision assisted absorption) is by far the best mechanism to generate hot carriers. The average energy of hot carriers is $\hbar\omega/2$ and their angular distribution is relatively directional (factor of 4). Most important, the carriers are all generated right at the surface and thus are capable of being emitted provided they have enough kinetic energy in the forward direction. The rate of Landau damping surpasses the rate of phonon and defect assisted damping when the mode penetration (or the entire nanoparticle size) becomes less than mean free path.

The recommendation is than very simple: one should use metal structures with dimensions less than mean free path, so that Landau damping will dominate the absorption process as long as the wavelength is long enough to avoid interband absorption and absorption due to electron-electron scattering.

Hopefully, this physically transparent treatment of all the processes of hot carrier generation by plasmons, will be useful to the developers of practical devices.